# Detection of Faults in Power System Using Wavelet Transform and Independent Component Analysis

[1]Prakash K. Ray, [2]B. K. Panigrahi, [2]P. K. Rout

[1]*Dept. of Electrical and Electronics Engineering, IIIT, Bhubaneswar, India.*

[2] *Dept. of Electrical Engineering, SOA University, Bhubaneswar, India.*

[3]Asit Mohanty, [4]Harishchandra Dubey

[3] *Dept. of Electrical Engineering, CET, Bhubaneswar, India.*

[4] *Dept. of ECE, The University of Texas at Dallas, USA.*

ABSTRACT: Uninterruptible power supply is the main motive of power utility companies that motivate them for identifying and locating the different types of faults as quickly as possible to protect the power system prevent complete power black outs using intelligent techniques. Thus, the present research work presents a novel method for detection of fault disturbances based on Wavelet Transform (WT) and Independent Component Analysis (ICA). The voltage signal is taken offline under fault conditions and is being processed through wavelet and ICA for detection. The time-frequency resolution from WT transform detects the fault initiation instant in the signal. Again, a performance index is calculated from independent component analysis under fault condition which is used to detect the fault disturbance in the voltage signal. The proposed approach is tested to be robust enough under various operating scenarios like without noise, with 20-dB noise and variation in frequency. Further, the detection study is carried out using a performance index, energy content, by applying the existing Fourier transform (FT), short time Fourier transform (STFT) and the proposed wavelet transform. Fault disturbances are detected if the energy calculated in each scenario is greater than the corresponding threshold value. The fault detection study is simulated in MATLAB/Simulink for a typical power system.

## 1 INTRODUCTION

Modern power utilities require a efficient protection scheme to protect the system itself as well as the connected equipments for improving better performance under normal as well as abnormal/faulty operating scenarios. Now-a-days the electromechanical relays are replaced by digital relays because of their characteristics like faster operation, accuracy and reliability. The fault detection through digital relays and fault detector (FD) is very vital for implementing any real-time solutions (Phadke 1988, Dash 2000).

In the literature, so may work are for monitoring and identification of faults in power system. Fault situation can be identified by comparing the difference between the value between two consecutive cycles when higher than a threshold value based on Phasor (Sidhu et al. 2002, Sachdev et al., 1991). But, its demerit being the modeling of fault resistance. Kalman filter is used for fault detection based on estimation method (Chowdhury et al. 1991, Zadeh et al. 2010, & Girgis 1982). Wavelet transform (WT) is being used for the detection of fault disturbances in power system (Ukil & R. Živanović 2007) which detects the changes in the signal. Then, adaptive filters and wavelet transform in combination was used for fault identification in power system (Ray et al. 2010). But, these techniques affected by variation in frequency, noise etc.

This paper proposes an algorithm for fault detection using wavelet transform and independent component analysis. The sudden changes are to be detected by the detection techniques so that suitable solutions can be adapted to protect the power system from any type of disturbances. The proposed method performances are being analyzed under different operating scenarios like in presence of noise, harmonics and frequency variations. Wavelet transform (WT) and independent component analysis (ICA) are suitable candidates for detection of fault disturbances because of efficient time-frequency resolutions and their reliability to extract the suitable features for identification purpose. Again, the techniques are considered under different operating conditions in order to assess the robustness and accuracies (Pradhan et al. 2005).

The techniques used for detection analysis is presented in Section 2 followed by the algorithm of implementation in Section 3. Then, Section 4 explains the result analysis both qualitatively and quantitatively, followed by the conclusions in Section 5.

## 2 TECHNIQUES FOR FAULT DETECTION

This section describes the techniques used for identification of different types of faults in wind system connected to grid power system. The WT and ICA as detection techniques are presented briefly with their mathematical modeling. Different operating scenarios are taken as case studies to test these methods for fault identifications which help in assessing both normal as well as faulty operating conditions. The details of these methods are as follows.

### 2.1 *Wavelet transforms (WT)*

The wavelet transform is a signal processing algorithm which is useful in detection of abnormal operating conditions based on decomposition of the power signals into different ranges of frequencies by the help of a series of low-pass and high-pass filters. This usually provides us a time-frequency multi-resolution analysis that greatly useful for identifying any short of abrupt variations in the electrical parameters such as voltage, phase, current, frequency etc. Here, Daubechies4 (dB4) is being used as the mother wavelet basis function for the fault detection analysis (Ukil & R. Živanović 2007). Usually, the signal is divided into a set of approximate (a) and detail (d) co-efficient representing the low-frequency and high frequency bands respectively. The decomposition is as presented below in Figure 1.

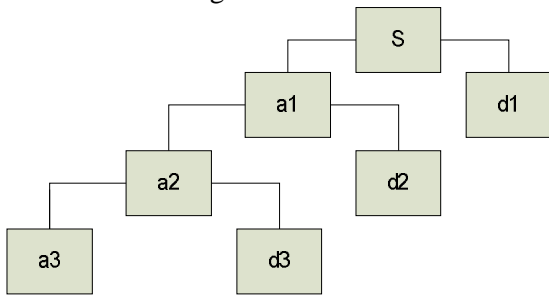

Figure 1: Wavelet decomposition tree

Considering a voltage signal of the power system as $V(t)$, the continuous wavelet transform (CWT) is expressed as:

$$CWT(V, M, N) = \frac{1}{\sqrt{a}} \int_{-\infty}^{\infty} V(t) \Psi^* \left( \frac{t-N}{M} \right) dt \quad (1)$$

Where $M$ and $N$ are called as dilation and translation parameters and $\Psi$ is known as the wavelet basis function. Now WT in discrete form as:

$$DWT(V, M, N) = \frac{1}{\sqrt{M_0^m}} \sum_k V(k) \Psi^* \left( \frac{n - kM_0^m}{M_0^m} \right) \quad (2)$$

Where M and N are replaced by $M_0^m$ and $kM_0^m$, and k, m are integers. Scaling function in one stage is expressed as sum of that of next stage and can be given by:

$$\varphi(t) = \sum_{n=-\infty}^{\infty} h(n) \sqrt{2} \; \varphi(2t - n) \quad (3)$$

Using the above equation, the original voltage signal can be written as [8]:

$$V(t) = \sum_{k=-\infty}^{\infty} a_{j_0}(k) 2^{j_0/2} \phi(2^{j_0} t - k) + \sum_{k=-\infty}^{\infty} \sum_{j=j_0}^{\infty} d_j(k) 2^{j/2} \phi(2^j t - k) \quad (4)$$

Where $j_0$ is the coarse adjustment parameter in the scaling function. The detail and approximate co-efficients are written:

$$a_j(k) = \sum_{m=-\infty}^{\infty} a_{j+1}(m) h(m - 2k) \quad (5)$$

$$d_j(k) = \sum_{m=-\infty}^{\infty} c_{j+1}(m) h_1(m - 2k) \quad (6)$$

### 2.2 *Independent Component Analysis (ICA)*

Independent Component Analysis (ICA) is an advanced or modified version of principal component analysis (PCA). It uses high order statistical analysis based de-correlation for the source input signal and provides important information regarding the irregularities presence. In ICA, the input data is modeled as linear coefficients that are mutually independent to each other. Based on the blind source separation algorithm, it can efficiently and accurately transform the input voltage signal into mutually and statistically independent components, thereby helps the detection process (Hyvärinen et al. 2001, Pöyhönen et al. 2003, Lopez et al. 2011, & Dubey et al. 2011).

The mutual information nothing but the individual independency measurement of the input signal and its entropy is Gaussian in nature and can be written as:

$$J(y) = h(v_{gaussian}) - h(v) \quad (7)$$

Differential entropy H of an input signal y with density $p_v(\eta)$ is given by:

$$h(v) = -\int p_h(\eta) \log p_v(\eta) d\eta \quad (8)$$

Here, the negentropy is estimated based on the estimation of probability functions of input signals Then, rather the negentropy can be expressed as:

$$J(v_i) = J\left(e\left(w_i^T x\right)\right) = \left[e\left\{g\left(w_i^T x\right)\right\} - e\left\{g\left(v_{gaussian}\right)\right\}\right]^2 \quad (9)$$

Here, e is the statistical expectation and G is a non-quadratic factor. For n linear mixtures taken as $x_1, x_2, \ldots x_n$ for the n independent components, we can express :

$$x_j = a_1 s_1 + a_2 s_2 + \ldots + a_n s_n \quad \text{For all } j \quad (10)$$

Here, $x_j$ is a random mixture and $s_k$ is a random independent variable. Assuming x as the random vector with elements $x_1, x_2, \ldots x_n$ and random vector with elements $s_1, s_2, \ldots s_n$, if A is the matrix having elements $a_{ij}$, we can take column vectors; $x^T$, as the transpose of x. Then, we can write in matrix form:

$$x = As \quad (11)$$

Where, A is a column matrix of elements a j and then model is expressed as:

$$x = \sum_{i=1}^{n} a_i s_i \quad (12)$$

The ICA mathematical model is an iterative one where the independent components are called latent variables. Here the input mix matrix is considered to be unknown and the components $s_i$ are statistically independent. Finally, once the matrix A is estimated, we can determine the inverse, W, then can evaluate the independent component by the following expression:

$$s = Wx \quad (13)$$

## 3 FAULT DETECTION METHODOLOGY

The fault detection methodology is explained in this section for the grid-connected wind power system as follows:
(i) A grid-connected wind power system is simulated in MATLAB where different types of fault are created. The voltage signal is extracted at PCC under faulty conditions. The signal is then passed through wavelet transform to get the time-frequency analysis for identification of faulty situations.
(ii) The voltage signal is processed through to find the mean and de-correlation. This is the 1st level of processing. .
(iii) Then, X data size is reduced and filtered for better data redundancy.
(iv) Then, the principal components (PC) of the input data are determined.
(v) In the current study, the independent components are calculated based on fixed point iteration [12]-[15].
(vi) Then, de-correlating matrix, $W_f$ and the independent component, $S_f$ of voltage signal is determined from which a matrix, $x_f$ is generated.
(vii) And finally, the pre-fault signal is considered for constructing the matrix, $x_n$ which can be used to calculate the performance index and fault is detected when the index is more than a set threshold value. The performance index is given by

$$PI(k) = (normal(absolute(W_f(k)*x_n(k) - s_f(k))))^2) \quad (14)$$

## 4 SIMULATED RESULTS

This section explains the detection results and corresponding discussions in a wind system connected to grid. The power system is simulated in MATLAB/Simulink environment and the grid is of 230 kV, 50 Hz rating consisting of thermal based power plant. The considered grid-connected power system is shown in Figure 2 and here, the sampling frequency is taken as 2 kHz.

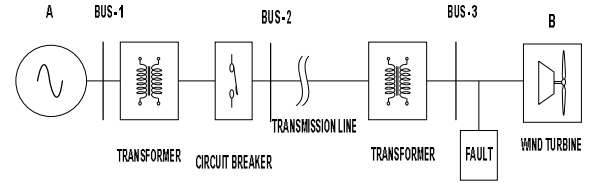

Figure 2: Grid-connected wind power System.

### 4.1 *Fault detection using wavelet transform*

The voltage signal at point of common coupling (PCC) is taken offline under AG fault. The signal is processed through WT which detects the fault instant based on filtering through a set of low-pass and high-pass filters. The voltage signal with its detection result is shown in Figure 3. Similarly, the detection result using AB and ABCG fault is shown in Figure 4 and Figure 5 respectively. These results clearly show that WT nicely detect the fault initiation instant nicely. But, as noticed when the fault is cleared, as shown in Figs., the recovery to normal waveform is un-noticed by wavelet transform.

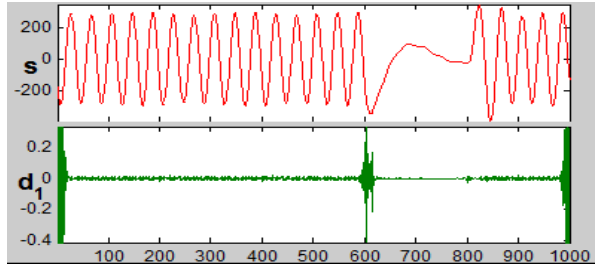

Figure 3: Detection of phase-to-ground fault in wind system connected to grid (red- voltage signal; green-WT output)

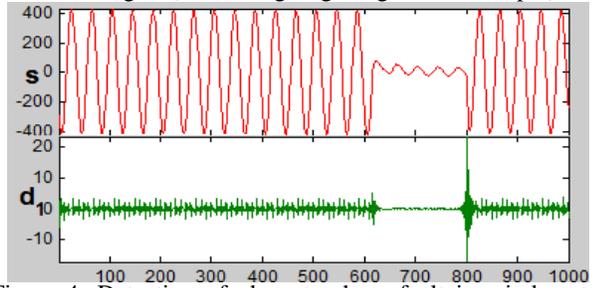

Figure 4: Detection of phase-to-phase fault in wind system connected to grid (red- voltage signal; green-WT output)

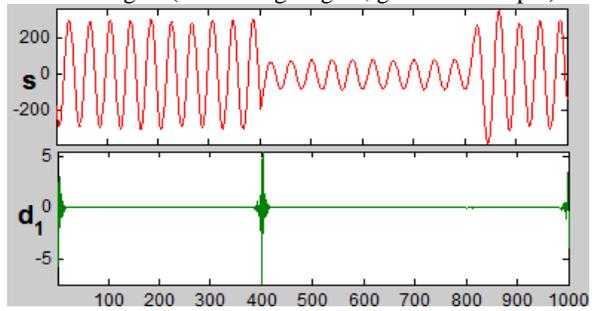

Figure 5: Detection of three-phase-to-ground fault in wind system connected to grid (red- voltage signal; green-WT output)

### 4.2 Phase-to-ground fault detection using Independent Component Analysis

A phase-to-ground fault is near to the point of common coupling and the corresponding voltage signal at the PCC is taken offline. Then, the index as explained in ICA section is calculated under different operating conditions like without noise, with 20-dB noise and under variation in frequency condition. The simulated of the performance index obtained from ICA is shown in Figure 6 (a). It is observed that, the fault initiated at near about 0.065 s detected through sudden increase in the index value. Before the fault, the index value is observed to be almost zero. Once, the fault occurs, the value increases suddenly and if we set a threshold value, then the can be detected. But in this case, we have not set the threshold value. Rather, we are assessing the faulty condition based on sudden increase.

### 4.3 Phase-to-phase fault detection using Independent Component Analysis

Next, a phase-to-phase fault is near to the point of common coupling and the corresponding voltage signal at the PCC. Then, the index as explained in ICA section is calculated under different operating conditions like without noise, with 20-dB noise and under variation in frequency condition. The simulated of the performance index obtained from ICA is shown in Figure 6 (b). It is observed that, the fault initiated at near about 0.065 s detected through sudden increase in the index value. Before the fault, the index value is observed to be almost zero. Once, the fault occurs, the value increases suddenly and if we set a threshold value, then the can be detected. But in this case, we have not set the threshold value. Rather, we are assessing the faulty condition based on sudden increase.

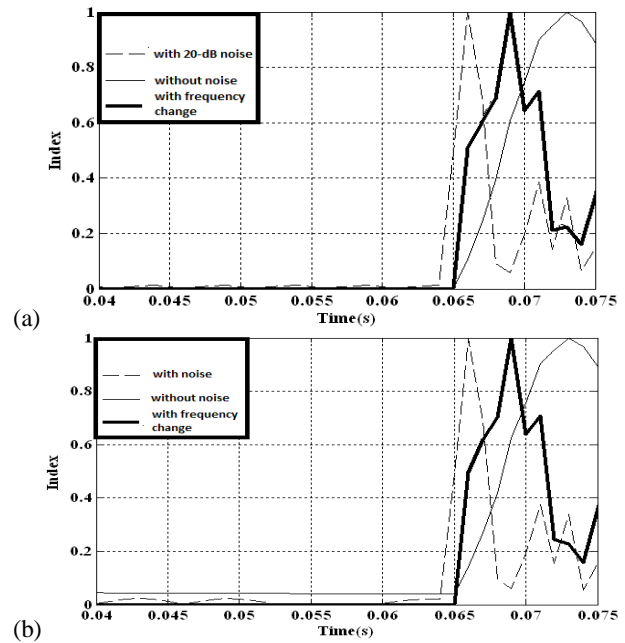

Figure 6: Detection of (a) phase-to-ground fault (b) phase-to-phase fault in wind system connected to grid

### 4.4 Detection of fault disturbances using the performance index (PI)

This sub-section describes the detection technique using a performance index called energy content of the processed faulty signal. The voltage signal at the point of common coupling is extracted and processed through different signal processing transforms like Fourier transform (FT), Short time Fourier transform (STFT) and wavelet transform. Then the output waveform after processing is used to calculate energy which is compared with a threshold value to know the faulty or normal operating conditions in the power system. It is observed that proposed WT provides larger energy value so that a suitable threshold can be chosen to accurately detect the faulty conditions. The index in different cases is shown in Table 1.

Table 1 Energy content using FT, STFT,WT

| Scenario | Energy content | | |
|---|---|---|---|
| | FT | STFT | WT |
| AG Fault | 1.3672 | 1.7863 | 2.1663 |
| BG Fault | 1.4525 | 2.2414 | 2.7352 |
| CG Fault | 1.3324 | 1.8367 | 2.6538 |
| AB Fault | 2.2341 | 2.3532 | 3.2514 |
| BC Fault | 2.6342 | 3.1230 | 3.8724 |
| ABC Fault | 3.1302 | 3.8225 | 4.2431 |

## 5 CONCLUSIONS

This paper presented detection of fault disturbances using Wavelet Transform and Independent Component Analysis. Voltage signal extracted at PCC is being processed through the above techniques under different fault and operating scenarios. It is observed that WT transform detected the fault instant very accurately. But, in some cases, it could not discriminate the disturbance condition may be because of noise or frequency variation. Therefore, the faults are accurately detected using ICA under all operating scenarios